\documentclass[a4 paper, 11pt]{article}
\pdfoutput=1
\usepackage{jheppub}

\usepackage{amsmath}
\usepackage{amssymb}
\usepackage{amscd}
\usepackage{dsfont}
\usepackage{enumerate}
\usepackage{amsfonts}
\usepackage{epsfig}
\usepackage{mathtools}
\usepackage{yfonts}
\usepackage{bbold}
\usepackage{breqn}
\usepackage[utf8]{inputenc}
\usepackage[english]{babel}
\usepackage{graphicx}
\usepackage{subcaption}

\def\bal#1\eal{\begin{align}#1\end{align}}
\def\alp[#1]{\begin{align}#1\end{align}}
\def\secnum[#1]{\texorpdfstring{$#1$}{TEXT}}
\def\secnuml#1\secnumr{\texorpdfstring{$#1$}{TEXT}}

\def\eqa{\begin{eqnarray}}
    \def\eqae{\end{eqnarray}}
\def\eq{\begin{equation}}
    \def\eqe{\end{equation}}
\def\be{\begin{equation}}
    \def\ee{\end{equation}}
\def\bea{\begin{eqnarray}}
    \def\eea{\end{eqnarray}}
\def\ba{\begin{array}}
    \def\ea{\end{array}}
\def\bd{\begin{displaymath}}
    \def\ed{\end{displaymath}}

\def\ie{{\it i.e.~}}

\def\>{\rangle}
\def\<{\langle}

\title{R\'enyi Entanglement of Purification and Half R\'enyi Reflected Entropy in Free Scalar Theory}

\author{Liangyu Chen} 
\affiliation{Yau Mathematical Sciences Center, Tsinghua University, Beijing 100084, China}

\emailAdd{liangyu-chen@mail.tsinghua.edu.cn}

\abstract
{In the AdS/CFT context, the entanglement of purification  (EoP, denoted as $E_{P}$) of CFT is conjectured to be dual to the entanglement wedge cross section (EWCS)  in bulk. However, another quantity called reflected entropy $S_{R}$ is also supposed to be dual to two times the EWCS. A natural question is whether they are the same in holographic CFTs even though they are different in general. Previous studies have shown $E_{P} \ge \frac{1}{2} S_{R}^{(n)}, n \ge2$ for random tensor networks. In this paper, we study this inequality beyond $n \ge 2$, and we focus on the range $0 < n < 2$. However, the calculations of EoP are notoriously difficult in general. Thus, our calculations mainly focus on the free scalar theory which is close to the holographic CFTs. We generalized the previous strategy for EoP in \cite{Takayanagi:2018sbw} to the R\'enyi case. And we have also presented two methods for R\'enyi reflected entropy, one is using correlators, the other one is Gaussian wavefunction ansatz. Our calculations show that the inequality still holds for $0 < n < 2$, and it may give us some insights into the equivalence of EoP and half reflected entropy in holographic CFTs. As byproducts of our research, we have also demonstrated the positivity of the R\'enyi Markov gap and the monotonicity of the R\'enyi reflected entropy in the free scalar theory. }

\begin{document}
\maketitle
    \section{Introduction}
    Understanding the quantum behavior of gravity is one of the most challenging problems in theoretical physics. Since its introduction by Juan Maldacena in 1997, the AdS/CFT correspondence has provided us with a non-perturbative approach for the study of quantum gravity \cite{Maldacena1999,Witten1998AntideSS,GUBSER1998105}. It states that a gravity theory in $d + 1$ dimensional Anti-de Sitter spacetime (AdS$_{d+1}$) is equivalent to a $d$ dimensional conformal field theory (CFT$_{d}$) on its boundary. While this statement is still a conjecture and there is no rigorous mathematical proof yet. Thus, theorists try to find more evidences to support it. There are many programs on it. One of the branches is using the entanglement structure to probe quantum gravity. In 2006, Ryu and Takayanagi had established the duality between the boundary entanglement entropy and the bulk minimal surface which ends on the boundary of the boundary subregion \cite{Ryu:2006bv, Ryu:2006ef}, this is the so-called RT formula.
    
    However, for mixed states, entanglement entropy is no longer a good measure for it. One needs to find other measures to quantify mixed states. From the CFT perspective, there are many candidates, such as mutual information, negativity, entanglement of purification (EoP) and reflected entropy etc. It is also very interesting to study the bulk dual of these measures. Recently, theorists made a conjecture that the EoP may have a very simple bulk dual. They conjectured that, based on some quantum information arguments, the bulk entanglement wedge cross section (EWCS) is dual to the EoP \cite{Takayanagi:2017knl, Nguyen:2017yqw}, there are many further developments on this duality \cite{Bao:2017nhh,  Tamaoka:2018ned, Hirai2018, Umemoto:2018jpc, Bao:2018gck,Espindola:2018ozt,Bao:2018fso,Bao:2018pvs,Caputa:2018xuf,BabaeiVelni:2019pkw,BabaeiVelni:2023cge}. After this conjecture, another quantity called reflected entropy was proposed as a kind of correlation measure. They also did the bulk calculations, and they showed that \cite{Dutta:2019gen}:
    \begin{equation}
        S_R(A: B)=2 E_W(A: B)=2 \frac{\operatorname{Area}\left[\Gamma_{AB}\right]}{4 G_N}+\ldots
    \end{equation}
    That is the half reflected entropy is also dual to the EWCS, see \cite{Jeong:2019xdr,Hayden:2021gno, Akers:2019gcv,Kusuki:2019evw, Moosa:2020vcs,Kudler-Flam:2020url,Boruch:2020wbe,Li:2020ceg,Bao:2019zqc,Chu:2019etd,Yuan:2024yfg} for more developments of it. Now, there are two quantities which both dual to EWCS, so one may ask whether the EoP equals the half reflected entropy in holographic CFTs at large N limit. Some progress has been made in this direction. Based on the random tensor networks, it has been shown that for any bipartite mixed state $\rho_{AB}$ and R\'enyi index $n \geq 2$, the following inequality holds \cite{Akers:2023obn}:
    \begin{equation} \label{mianIeq1}
        E_P(A: B) \geq \frac{1}{2} S_R^{(n)}(A: B) 
    \end{equation} 
    If one can generalize this inequality to the n = 1 case, then combine with the following inequality \cite{Akers:2023obn}:
    \begin{equation} \label{mianIeq2}
        E_P(A: B) \leq E_{W}(A: B) = {1 \over 2} S_R(A: B)
    \end{equation}
    which is true for Random Tensor Networks (RTNs) (Note that, this inequality is not trivial, since from the minimal definition of EoP, we can only conclude that $E_P(A: B) \leq S_R(A: B)$ instead of  $E_P(A: B) \leq {1\over 2} S_R(A: B)$ ). The key idea of this inequality is that they found, in RTNs, $E_{W}(A:B)$ can be approximately written as entanglement entropy of a global pure state, and we know $E_{P}$ is the minimal one of all these purifications, so  \eqref{mianIeq2} holds.
    
    Then with these two inequalities, one can show that the equality of the EoP and reflected entropy at the leading order. Previous work \cite{Couch:2023pav} showed that there is a counter example for the above inequality when $0 < n <2$. However, their model is not a holographic CFTs. Thus, the conjecture of holographic purification entropy is still possible to be true. In this paper, we study a generalized version of this inequality for $0 < n < 2$ in free scalar theory. We will consider the inequality:
    \begin{equation} \label{BasicInequality}
        E^{(n)}_P(A: B) \geq \frac{1}{2} S_R^{(n)}(A: B) 
    \end{equation} 
    Our results showed that \eqref{BasicInequality} still holds in this case. Due to the monotonicity of R\'enyi EoP, this inequality is stronger than \eqref{mianIeq1} for $n>1$ and weaker for $n<1$. Of course, one may want to know whether this is still hold in the case of true holographic CFTs. While, even the EoP is notoriously difficult to compute in holographic CFTs, let alone R\'enyi EoP. However, we would like to point out that even our model is free scalar, it may shed some light to support the equality between EoP and half reflected entropy.
    
    This paper is organized as follows. In Sec.\ref{sec:BasicReview}, we review the general aspects of EoP and reflected entropy. In Sec.\ref{sec:$E_{P}^{(n)}$}, we present our setup for free scalar theory, then we review the strategy developed in \cite{Takayanagi:2018sbw} for EoP in free scalar, after that we present the formula for R\'enyi EoP. In Sec.\ref{sec:RenyiReF}, we present two methods for R\'enyi Reflected entropy in free scalar theory. We give a pedagogical review for the correlator method of reflected developed in \cite{Bueno:2020fle}. With this knowledge and the key idea of correlator method for R\'enyi entanglement entropy \cite{Casini:2009sr}, one can write down the closed formula for R\'enyi reflected entropy directly. For more general discussions on $(m,n)$-R\'enyi reflected entropy $S_{m,n}^{R}$, see \cite{Berthiere:2023gkx}. We also present the Gaussian wave function method for R\'enyi reflected entropy in this section, this is because the reflected entropy is also a purification, we can apply the method for R\'enyi EoP on it with a few modifications.  In Sec.\ref{sec:NumericalResult}, we present our numerical results for R\'enyi EoP and half R\'enyi reflected entropy, we mainly focus on their difference for different R\'enyi index. The Sec.\ref{sec:Summary} is for summary and discussions. In App.\ref {app:ReFn}, we give more details on the review of correlator method to make the discussion self-contained.

   \section{Basic review for EoP and Reflected Entropy} \label{sec:BasicReview}
   In this section, we give a quick review for EoP and reflected entropy. The EoP is a kind of correlation measure between two subsystems and it is defined as the following. Consider a mixed state $\rho_{AB}$ between two subsystems  $ A $ and $ B $, then one can use two auxiliary systems $A^{\prime}, B^{\prime}$ to purify it. Let the purified state be $ |\psi\rangle_{AB A^{\prime} B^{\prime}}$, then by taking the partial trace, we get the reduced density matrix $ \rho_{AA^{\prime}} = \mathrm{Tr}_{BB^{\prime}} \left(|\psi \rangle  \langle \psi | \right)  $,  with this we get the  Von Neumann entropy:
   \begin{equation}
      S(\rho_{AA^{\prime}}) = -\mathrm{Tr} \left( \rho_{AA^{\prime}} \log \rho_{AA^{\prime}} \right)
   \end{equation}
   There are infinity ways to do the purification. The Entanglement of Purification (EoP) is defined as the minimum  of $S(\rho_{AA^{\prime}})$, that is \cite{Terhal:2002riz}:
   \begin{equation}
       E_{P}(A: B) =\operatorname{Min}_{|\psi \rangle_{ABA^{\prime}B^{\prime}} } S(\rho_{AA^{\prime}})
   \end{equation}
   Due to this minimization process , it is hard to calculate in general. However, under some reasonable assumptions, it is possible to calculate the EoP in free scalar\cite{Takayanagi:2018sbw}. In this paper, we will follow this and generalize it to R\'enyi case. We leave the details of calculations to the next section.
   
   The reflected entropy is the $S(\rho_{AA^{\prime}})$ by doing a canonical purification. Let $\rho_{AB}$ has the following spectrum decomposition:
   \begin{equation}
       \rho_{AB} = \sum_{i} \lambda_{i}|\lambda_{i} \rangle \langle \lambda_{i} |
   \end{equation} 
  the canonical purification is defined as reflecting the bra state to ket state (the so called Choi-Jamiolkowski isomorphism) and taking the square root of the eigenvalues.
  \begin{equation}
     |\psi \rangle_{AB A^{\prime} B^{\prime}} = |\sqrt{\rho_{AB}} \rangle \equiv  \sum_{i} \sqrt{\lambda_{i} }|\lambda_{i} \rangle |\lambda_{i} \rangle^{\prime}
  \end{equation}
   where the upper index "prime" means it is state in the double Hilbert space $H^{*}$. Of course, we can write the reflected state in general basis by doing a unitary transformation. In general, we have $  \rho_{AB} = \sum_{ab, \tilde{a}\tilde{b}} \rho_{ab,\tilde{a}\tilde{b}} |ab \rangle \langle \tilde{a}\tilde{b}|$, then the purified state is given by
   \begin{equation}
       | \sqrt{\rho_{AB}}  \rangle= \sum_{ab, \tilde{a}\tilde{b}} \sqrt{\rho_{ab,\tilde{a}\tilde{b}}} |ab \rangle |\tilde{a}\tilde{b}\rangle^{\prime}
   \end{equation}
   
   For field theory, it is hard to calculate the reflected entropy straightforwardly, since we don't know the spectrum in general like the calculation of entanglement entropy. Thus, usually the replica trick is used. Let us review the replica trick for reflected entropy which is a little bit different from the one of entanglement entropy. We start from the following state:
   \begin{equation}
       \left|\psi_m\right\rangle=\frac{1}{\sqrt{\operatorname{Tr} \rho_{A B}^m}}\left|\rho_{A B}^{m / 2}\right\rangle
   \end{equation} 
  The canonical purified state is the $ m = 1 $ case. Then the two indexes $(m,n)$-R\'enyi reflected entropy is defined by:
  \begin{equation}
      S^{(m, n)}_R(A:B)=\frac{1}{1-n} \log \operatorname{Tr}\left(\rho_{A A^{\prime}}^{(m)}\right)^n =\frac{1}{1-n} \ln \frac{Z_{n, m}}{\left(Z_{1, m}\right)^n}
  \end{equation}
  where the reflected density matrix $\rho_{A A^{\prime}}^{(m)}=\operatorname{Tr}_{BB^{\prime}}\left(\left|\psi_m\right\rangle\left \langle\psi_m\right|\right)$ and $Z_{n,m}$ is the partition:
  \begin{equation}
      Z_{n, m} \equiv \operatorname{Tr}_{A A^{\prime}}\left(\operatorname{Tr}_{B B^{\prime}}\left|\rho_{A B}^{m / 2}\right\rangle\left\langle\rho_{A B}^{m / 2}\right|\right)^n
  \end{equation}
  which can be calculated by using the Euclidean path integral in some cases. The usual R\'enyi reflected entropy is recovered by taking $m=1$. And the reflected entropy is obtained  by taking $n=1$ further.

  \subsection{The gravity dual: Entanglement Wedge Cross Section}
    \begin{figure}[t]
      \centering
      \includegraphics[width=0.4\linewidth]{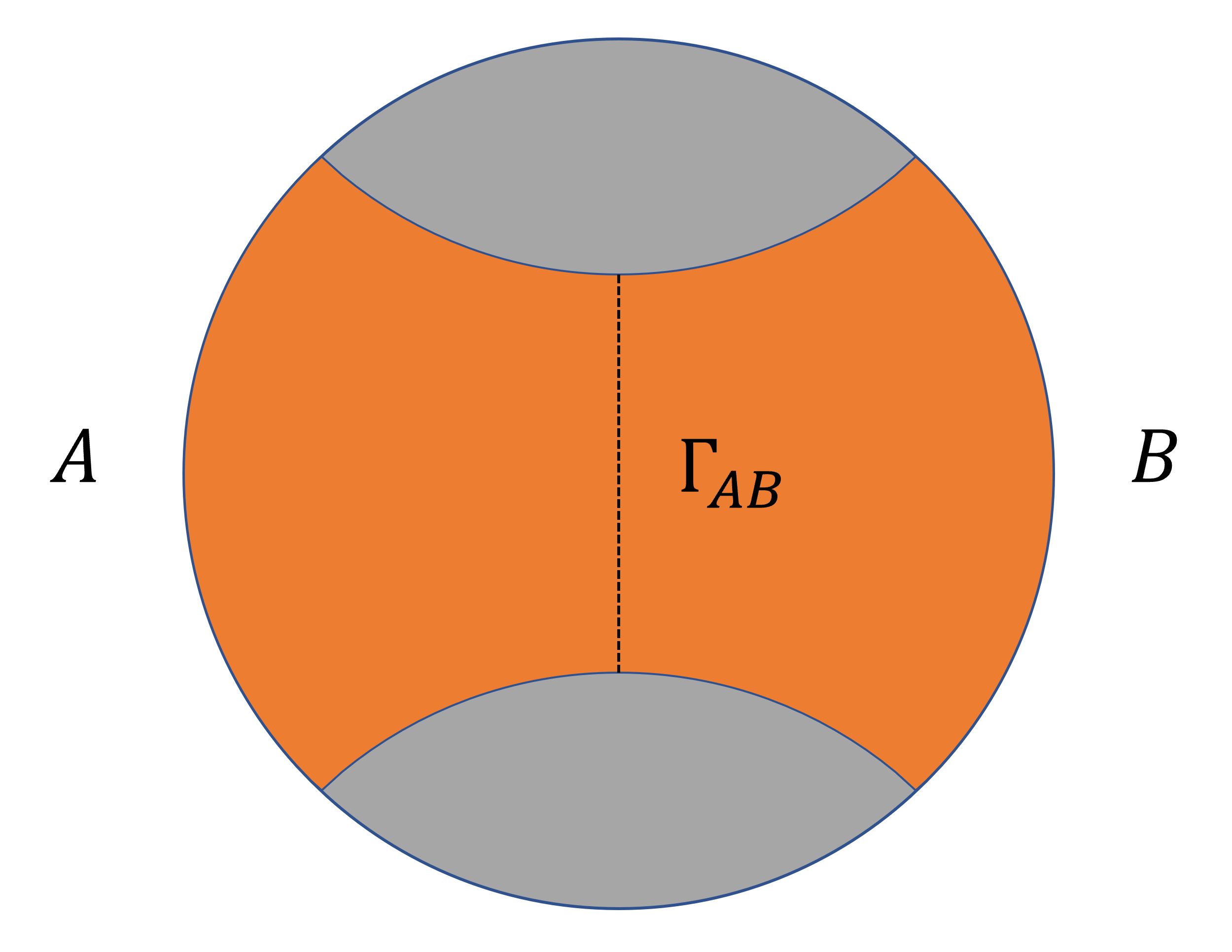}
      \caption{ This is a time slice of global AdS$_{3}$. $A,B$  are the two disjoint subregions in the boundary. The orange region is the entanglement wedge of $A \cup B$. The dashed line $\Gamma_{AB}$ is the minimal surface  which connect the two RT surfaces. The area of $\Gamma_{AB}$ divided by $4G_{N}$ is defined as entanglement wedge cross section (EWCS). }
      \label{fig:EWCS}
  \end{figure}
   In the spirit of AdS/CFT correspondence, it is natural to ask what is the gravity dual of EoP and reflected entropy, respectively. Now, let us discuss the so-called Entanglement Wedge Cross Section (EWCS) in the bulk. And see, how it relates to information measures in the boundary.

  To illustrate the EWCS,  we can consider the time slice of global AdS$_{3}$, which is a disk. Consider the two disjoint subregions $A, B$ in the boundary, see Fig. \eqref{fig:EWCS}. We will consider the situation when the mutual information of this setup is nonzero, \ie the RT surface of $A \cup B$ is not the union of A's and B's. Then, EWCS is defined as the area of minimal surface connecting the two disjoint parts of $A \cup B$'s RT surface.
  \begin{equation}
      E_{W} = \frac{\mathrm{Area}(\Gamma_{AB})}{4G_{N}}
  \end{equation}

  \subsubsection{Holographic Entanglement of Purification}
  There is a heuristic way to explain why $E_{P}$ is conjectured to equal $E_{W}$. One can remove the complement of $A \cup B$ firstly, then we have two disjoint part $A$ and $B$, after that we add two auxiliary part $\tilde{A}$ and $\tilde{B}$ to glue the previous disjoint part as a whole system $A\tilde{A}B\tilde{B}$. And this whole system is still corresponding to a pure state. One can imagine this is a new extended CFT, so the entanglement entropy $S_{A\tilde{A}}$ is dual to bulk RT surface $\Gamma_{AB}$. Since the $E_{P}$ is the minimal of $S_{A\tilde{A}}$, so the bulk of $E_{P}$ will be given by the EWCS.
    \begin{figure}[t]
      \centering
      \includegraphics[width=0.4\linewidth]{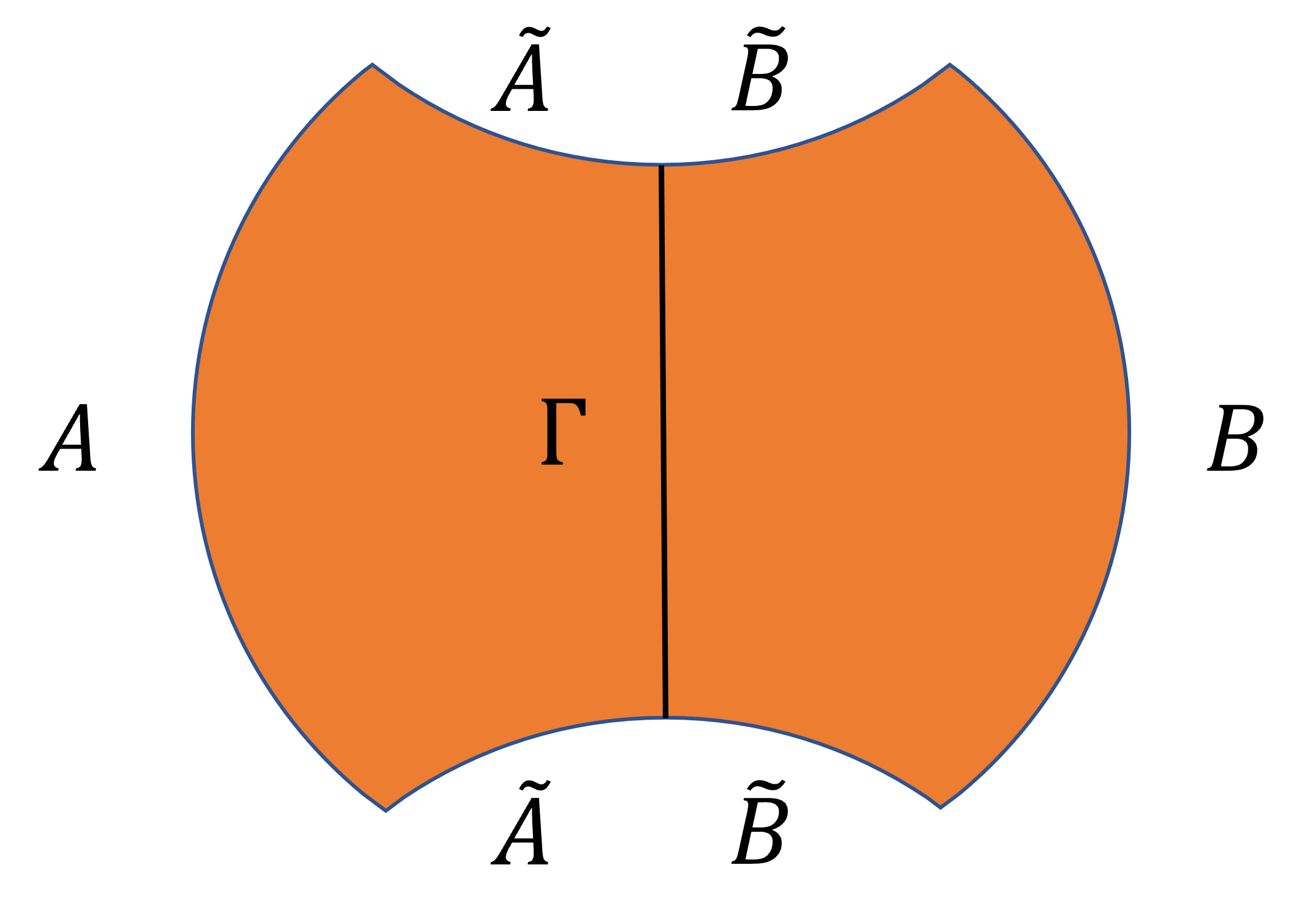}
      \caption{The illustration for holographic EoPs. This is the time slice of global AdS$_{3}$ with the complement of $A\cup B$  is removed and two auxiliary part $\tilde{A}$ and  $\tilde{B}$ are added. $\Gamma$ is the RT surface for subregion $A\tilde{A}$. }
      \label{fig:EWCSHeuristic}
  \end{figure}

  \subsubsection{Holographic Reflected entropy}
   After the holographic purification conjecture was proposed, Dutta and Faulkner studied the holographic dual of reflected entropy. And they found that the EWCS equals half reflected entropy at the leading order of $G_{N} $ \cite{Dutta:2019gen}
   \begin{equation}
    S_R(A:B)=2E_W(A:B)+ \mathcal{O}(G_{N}^{0}) 
   \end{equation}
   For this statement, there is also an intuitive explanation. Remember that the canonical purification is making a mirror for original Hilbert space. Thus, the reduced density matrix $\rho_{AB}$ and  $\rho_{\tilde{A} \tilde{B}}$ are the same. So, the total entanglement wedge is two copies of single one and gluing them alone one side of RT surface. Then reflected entropy is dual to the area minimal surface dividing $A\tilde{A}$ and $B\tilde{B}$.  Due to the symmetry between $AB$ and $\tilde{AB}$, this area will be twice of $E_{W}(A:B)$. 
  
  \section{The R\'enyi EoP in free scalar theory } \label{sec:$E_{P}^{(n)}$}
   Let us focus on R\'enyi EoP in free scalar theory in this section. We consider the same model as in literature \cite{Takayanagi:2018sbw}. In \cite{Takayanagi:2018sbw}, they proposed a Gaussian Ansatz to calculate $E_{P}$, but their calculations only focus on entanglement entropy. For our purpose, we need the R\'enyi entropy. So, in this section. We will show how to generalize their calculations to general R\'enyi case. We will give a closed formula for R\'enyi EoP in this section and leave the numerical result in Sec. \ref{sec:NumericalResult}

   \subsection{Discretization of free scalar}
  Let us focus on a two dimensional massive free scalar, the Hamiltonian is given by
   \begin{equation}
       H_0=\frac{1}{2} \int d x\left[\pi^2+\left(\partial_x \phi\right)^2+m^2 \phi^2\right] .
   \end{equation}
 After the  discretization procedure, the rescaled Hamiltonian (with periodic boundary condition imposed )becomes  
 \begin{equation}
    \label{model_H}
     H= a H_{0} = \sum_{n=1}^N \frac{1}{2} \pi_n^2+\sum_{n, n^{\prime}=1}^N \frac{1}{2} \phi_n V_{n n^{\prime}} \phi_{n^{\prime}}
 \end{equation}
 where $a$ is lattice spacing (we will also choose $a=1$ in numerical calculations ), $N$ is total number of this chain and
 \begin{equation}
     V_{n n^{\prime}}=N^{-1} \sum_{k=1}^N\left[a^2 m^2+2(1-\cos (2 \pi k / N))\right] e^{2 \pi i k\left(n-n^{\prime}\right) / N}.
 \end{equation}
 The ground state of this model is given by:
 \begin{equation}
     \begin{aligned}
          \Psi_0[\phi] &= \mathcal{N}_0 \cdot e^{-\frac{1}{2} \sum_{n, n^{\prime}=1}^N \phi_n W_{m n} \phi_n^{\prime}} \\
           W_{n n^{\prime}} &=\sqrt{V}_{n n^{\prime}}=\frac{1}{N} \sum_{k=1}^N \sqrt{a^2 m^2+2(1-\cos (2 \pi k / N))} e^{2 \pi i k\left(n-n^{\prime}\right) / N},
     \end{aligned}
 \end{equation}
  where $\mathcal{N}_{0}$ is normalization constant.  And $W$ matrix will play a key role as an input in the following discussions.
   
  \subsection{The EoP  in  Gaussian Ansatz}
   To compute the EoP, we need to choose a global pure state, usually we choose the ground state. By tracing out the complement of $A \cup B$, we obtain the target mixed state which we want to purify. Then, one needs to consider all possible ways for purification to get the minimal entanglement entropy $S_{A\tilde{A}}$
   
  Let the subsystems are two disjoint subregion $A$ and $B$, then the ground state can be rewritten as :
    \begin{equation}
        \Psi_0\left[\phi_{A B}, \phi_{A_{0} B_{0}}\right]=\mathcal{N}_0 \cdot \exp \left[-\frac{1}{2}\left(\phi_{A B}, \phi_{A_{0} B_{0}}\right)\left(\begin{array}{cc}
            J_0 & K_0 \\
            K_0^T & L_0
        \end{array}\right)\left(\begin{array}{c}
            \phi_{A B} \\
            \phi_{A_{0} B_{0}}
        \end{array}\right)\right]
        \label{GroundState}
    \end{equation}
   where $\phi_{AB} $ means the components of $\phi$ with index in subregion $A \cup B$, and $A_{0} B_{0}$ is the complement of $AB$. After taking the partial trace, we get the reduced density matrix:
   \begin{equation}
       \begin{aligned}
           & \rho_{A B}\left[\phi_{A B}, \phi_{A B}^{\prime}\right] \\
           &= \mathcal{N}_{AB} \cdot \exp \left[-\frac{1}{2}\left(\phi_{A B}, \phi_{A B}^{\prime}\right)\left(\begin{array}{cc}
               J_0 + M_0 &  M_0 \\
                M_0 & J_0 +M_0
           \end{array}\right)\left(\begin{array}{c}
               \phi_{A B} \\
               \phi_{A B}^{\prime}
           \end{array}\right)\right] .
       \end{aligned}
   \end{equation}
where $M_0 = \frac{1}{2} K_0 L_0^{-1} K_0^T$. 

To purify $\rho_{AB}$, we will use the Gaussian ansatz, \ie we will assume the purified state is still Gaussian. Then, we can still use the above structure to represent the purified state:
\begin{equation}
    \Psi\left[\phi_{A B}, \phi_{\tilde{A}\tilde{B}}\right]=\mathcal{N}_0 \cdot \exp \left[-\frac{1}{2}\left(\phi_{A B}, \phi_{\tilde{A}\tilde{B}} \right)\left(\begin{array}{cc}
        J & K \\
        K^T & L
    \end{array}\right)\left(\begin{array}{c}
        \phi_{A B} \\
       \phi_{\tilde{A}\tilde{B}}
    \end{array}\right)\right]
\end{equation}
 with the following constraint equations:
 \begin{equation} \label{purifiedConstraint}
     J= J_0, \quad M = M_{0}
 \end{equation}
 where $\tilde{A}\tilde{B}$ is the auxiliary system and  $M \equiv \frac{1}{2} K L^{-1} K^T $.
 
  To obtain the EoP, we only need to parameterize these matrices under these constraints and find the one which minimize the entropy $S(\rho_{A A^{\prime}})$. For the R\'enyi EoP which we will discuss soon, we just replace the entropy $S(\rho_{A A^{\prime}})$ with the corresponding  R\'enyi generalization.
 
 For these Gaussian states, there is a standard formula to calculate the entropy.  Consider a pure state $|\psi \rangle_{AB} $ of the total system $AB$. Since it is a Gaussian state, we can use the following ansatz:
  \begin{equation}
      \label{tot-pure-func}
      \Psi_{A B}=\mathcal{N}_{A B} \cdot \exp \left[-\frac{1}{2}\left(\phi_A,  \phi_B\right)\left(\begin{array}{cc}
          A & B \\
          B^T & C
      \end{array}\right)\left(\begin{array}{c}
          \phi_A \\
          \phi_B
      \end{array}\right)\right]
  \end{equation}
  Let 	
  \begin{equation}
      W=\left(\begin{array}{cc}
          A & B \\
          B^T & C
      \end{array}\right), \quad W^{-1}=\left(\begin{array}{cc}
          D & E \\
          E^T & F
      \end{array}\right)
  \end{equation}
  Then, the entanglement entropy between $A$ and $B$ is given by \cite{Bombelli1986}
  \begin{equation}
      S_{A} = S_{B} = \sum_{i =1}^{|A|} f(\lambda_{i}),
      \label{entropy-formula}
  \end{equation}
  where $ \lambda_{i} $ is the eigenvalue of the  matrix $ \Lambda = -E\cdot B^{T} = D \cdot A  - I $.  And the function $ f(x) $ is given by:
  \begin{equation}
      f(x)=\log \frac{\sqrt{x}}{2}+\sqrt{1+x} \log \left(\frac{1}{\sqrt{x}}+\frac{\sqrt{1+x}}{\sqrt{x}}\right)
      \label{formula-Gaussian-entropy}
  \end{equation} 
 For our situation, we care about entanglement entropy between $A\tilde{A}$ and $B \tilde{B}$. So, to calculate the EoP, we need to rearrange the block matrix:
 \begin{equation}
     \left(\begin{array}{cc}
        J & K \\
        K^T & L
    \end{array}\right)
 \end{equation}
 into a new block matrix with partition $A\tilde{A}|B\tilde{B}$.
 \begin{equation}
    \hat{W} = \left(\begin{array}{cc}
        \hat{J} & \hat{K} \\
        \hat{K}^{T} & \hat{L}
    \end{array}\right)
 \end{equation}
 Taking its inverse:
 \begin{equation}
    \hat{W}^{-1} = \left(\begin{array}{cc}
        \hat{P} & \hat{R} \\
        \hat{R}^{T} & \hat{Q}
    \end{array}\right)
 \end{equation}
 Then, the EoP is given by:
 \begin{equation} \label{EoPLambda}
     E_{P}(A:B) =  \min S_{A\tilde{A}} = \min \sum_{i =1}^{|A\tilde{A}|} f(\hat{\lambda}_{i})
 \end{equation}
 where $\hat{\lambda}_{i}$ is the eigenvalue of matrix $\hat{\Lambda} = \hat{P} \hat{J} - I$.

 \subsection{Generalize to R\'enyi EoP}
 To show the inequality between R\'enyi EoP and Half R\'enyi reflected entropy, we need to generalize the above method to general R\'enyi index case. So, we need to know how \eqref{entropy-formula} is derived. And following its spirit,  generalizing it to R\'enyi case. The key point is establishing the spectrum relation between $\rho_{A}$ and $\Lambda$.  Then using it to calculate $\mathrm{Tr}(\rho_{A})^{n}$.
 
 Let us consider a system of two oscillators at first, each with one degree of freedom. Let $a$ and $b$ be the annihilation operators for the two oscillators. Consider the state vector
 \begin{equation}
     \begin{aligned}
         |\psi \rangle &= (1 - \gamma^{2})^{1/2} e^{\gamma a^{\dag} b^{\dag}} |0\rangle_{a} |0\rangle_{b} \\
         &=  (1 - \gamma^{2})^{1/2} \sum_{n=0}^{\infty} \gamma^{n} |n\rangle_{a} |n\rangle_{b} \\
     \end{aligned} 
 \end{equation}
 where $\gamma$ is a real number. Obviously, after tracing out the $b$ part, the reduced density matrix becomes 
 \begin{equation}
     \begin{aligned}
         \rho_{a} &= \mathrm{Tr}_{b} \left( | \psi \rangle \langle \psi | \right) =  (1 - \gamma^{2}) \sum_{n=0}^{\infty} \gamma^{2n} |n\rangle_{a} \langle n |_{a} \\
     \end{aligned} 
 \end{equation}
 Then, the entropy becomes 
 \begin{equation}
     \begin{aligned}
         S_{a} &= - \mathrm{Tr}  \left( \rho_{a} \ln \rho_{a}\right) = -  \ln (1- \gamma^{2})  - \frac{\gamma^{2}}{1 - \gamma^{2}} \ln \gamma^{2}
     \end{aligned} 
 \end{equation}
 Similarly the R\'enyi entropy becomes
 \begin{equation}
     \begin{aligned}
         S_{a}^{(n)} &= \frac{1}{1-n} \ln  \mathrm{Tr} \rho_{a}^{n} = \frac{1}{1-n} \ln \frac{(1 - \gamma^{2})^{n}}{ 1- \gamma^{2n}}
     \end{aligned} 
 \end{equation}
 When the state is generalized to multi-oscillators state like \eqref{tot-pure-func},  then the theorem in \cite{Bombelli1986} tells us, the entropy is given by 
 \begin{equation}
     S_{A}  = \sum_{i =1}^{|A|} f(\lambda_{i}),  \quad f(x)=\log \frac{\sqrt{x}}{2}+\sqrt{1+x} \log \left(\frac{1}{\sqrt{x}}+\frac{\sqrt{1+x}}{\sqrt{x}}\right),
     \label{formula-Gaussian-entropy}
 \end{equation}
 and the relation is 
  \begin{equation}
     \label{gamma-eigvals-relation}
     \gamma^{2}(x) = \frac{x + 2 - 2 \sqrt{1+x}}{x}
 \end{equation}
 Then, by substituting it back to R\'enyi entropy, we get
 \begin{equation}
     S_{A}^{(n)} = \sum_{i =1}^{|A|} f^{(n)}(\lambda_{i}), \quad \text{with} \;\ f^{(n)}(x) = \frac{1}{1-n} \ln \frac{(1 - \mu(x))^{n}}{1- \mu^{n}(x) }
 \end{equation}
 where we denote $\mu(x) \equiv \gamma^{2}(x)$. As one can directly check, $f^{(n)}(x)$ will reduce to $f(x)$ when $n=1$. For this case, the analytical form of $f^{(n)}(x)$ is a little complicated. Numerically, we will try to get the eigenvalue first and then substitute into \eqref{gamma-eigvals-relation} to calculate the R\'enyi entropy.
 
 Now, returning back R\'enyi EoP,  it is just given by:
  \begin{equation}
     E_{P}^{(n)}(A:B)= \min S_{A\tilde{A}}^{(n)} =  \min \sum_{i =1}^{|A\tilde{A}|} f^{(n)}(\hat{\lambda_{i}})
 \end{equation}
 With this formula in hand, we can compute R\'enyi EoP systematically. We leave the numerical result in Sec. \ref{sec:NumericalResult}

 \section{The R\'enyi reflected entropy in free scalar theory} \label{sec:RenyiReF}
  R\'enyi reflected entropy is an another important ingredient for the inequality \eqref{BasicInequality}. In this section, we will present two methods to calculate the reflected entropy for free scalar, then we will show 
 how to generalize it to R\'enyi case.

 \subsection{Correlator method for R\'enyi reflected entropy }
 Now, let us review the method developed by Bueno and Casini for the calculation of reflected entropy \cite{Bueno:2020fle}. Remember that the Hamiltonian of our model in the lattice is
 \begin{equation}
     H=  \sum_{n=1}^N \frac{1}{2} \pi_n^2+\sum_{n, n^{\prime}=1}^N \frac{1}{2} \phi_n V_{n n^{\prime}} \phi_{n^{\prime}}
 \end{equation}
 The key point of their method is using the correlator to represent the reflected entropy. To compute the reflected entropy, the first step is extending the original Hilbert space $\mathcal{H}_{AB}$ of sub-systems $A,B$ with a double copy $\mathcal{H} = \mathcal{H}_{AB} \otimes \mathcal{H}_{\tilde{A} \tilde{B}}$. Then, the field operator $\phi_{i}$ and momentum operator $\pi_{j}$, $i,j = 1, 2, \cdots N$, are extended by the following modular conjugation 
 \begin{equation}
     \tilde{\phi}_{i} = J \phi_{i} J, \quad  \tilde{\pi}_{i} = -J \pi_{i} J
 \end{equation}
 one can check these $4N$ operators form the canonical commutation relations. Now, let us arrange these operators in tho $\Phi$ and $\Pi$, s.t.
 \begin{equation}
  \begin{aligned}
      \Phi_{i} &= \phi_{i},  \quad \Phi_{i+N} &= \tilde{\phi}_{i} \\
       \Pi_{i} &= \pi_{i},  \quad \Pi_{i+N} &= \tilde{\pi}_{i} 
  \end{aligned}
 \end{equation}
 where $i=1,2, \cdots N $. We will focus on the following two correlators with the index $i= 1,2, \cdots, 2N$ now:
 \begin{equation}
     G_{ij}^{\Phi} = \langle \Omega | \Phi_{i} \Phi_{j} |\Omega \rangle, \quad G_{ij}^{\Pi} = \langle \Omega | \Pi_{i} \Pi_{j} |\Omega \rangle
 \end{equation}

With all these ingredients in hand, we can obtain the sub-matrix of $G^{\Phi}$ and $G^{\Pi}$ with indexes support in the subregion $ A \cup \tilde{A} $, then one can  diagonal the following matrix:
\begin{equation}
    C_{A\tilde{A}} = \sqrt{G^{\Phi}_{A\tilde{A}} G^{\Pi}_{A\tilde{A}}}
\end{equation}
Let the eigenvalue of this matrix is $v_{k}$, then one can derive the following relations:
\begin{equation}
    v_{k} = \frac{1}{2} \coth (\epsilon_k/2) 
\end{equation}
where ${\epsilon}_{k}$ is the eigenvalue of modular Hamiltonian $H_{A\tilde{A}}$. Then the reflected entropy is given by:
\begin{equation}
\begin{aligned}
    S_{R}(A : B)= S_{Von}(\rho_{A \tilde{A}}) &= \operatorname{tr}\left[\left( C_{A\tilde{A}}+1 / 2\right) \log \left(  C_{A\tilde{A}} + 1 / 2\right)-\left( C_{A\tilde{A}}-1 / 2\right) \log \left(  C_{A\tilde{A}}-1 / 2\right)\right] \\
    &= \sum_{k} \left[\left( v_{k}+1 / 2\right) \log \left(  v_{k} + 1 / 2\right)-\left( v_{k}-1 / 2\right) \log \left( v_{k} -1 / 2\right)\right]
\end{aligned}
\label{RefE}
\end{equation}
where the index $k$ is taken from $A \tilde{A} $.

From the above procedure, we know in order to compute the reflected entropy, we need to know the correlators $G^{\Phi}$ and $G^{\Phi}$ in doubled Hilbert space $\mathcal{H}$. For the more common entanglement entropy, one just need the correlator in the original Hilbert space, and this is the main difference between the correlator method for reflected entropy and that for entanglement entropy. It was shown that  $G^{\Phi}$ and $G^{\Pi}$ can be represented as:
\begin{equation}
    G^{\Phi}=\left(\begin{array}{cc}
        X & g(X P) X \\
        g(X P) X & X
    \end{array}\right), \quad G^{\Pi}=\left(\begin{array}{cc}
        P & -P g(X P) \\
        -P g(X P) & P
    \end{array}\right)
\end{equation}
where $g(A) = (A - 1/4)^{1/2} A^{-1/2}$ and
  \begin{equation}
        X_{i j}  =\left\langle\phi_i \phi_j\right\rangle, \qquad
        P_{i j}  =\left\langle\pi_i \pi_j\right\rangle
\end{equation}
are the correlators in original Hilbert space $\mathcal{H}_{AB}$. For the our concrete model here, it can be shown further that they are given by
\begin{equation}
     X_{i j}= \frac{1}{2}\left(V^{-\frac{1}{2}}\right)_{i j} = \frac{1}{2} W^{-1}_{ij}, \qquad P_{i j} = \frac{1}{2}\left(V^{\frac{1}{2}}\right)_{i j} = \frac{1}{2} W_{ij}
\end{equation}

\subsubsection*{Renyi reflected entropy}
In this paper, we also need to compute the R\'enyi reflected entropy. In general, we will use the replica trick to convert the  R\'enyi reflected entropy into Euclidean path integral.  Since we are considering free scalars, the thing is much simpler. We can continue to use the correlator method. The formula is given by:
 \begin{equation} \label{RenyiReF}
    S_{R}^{(n)}(A : B) = \frac{1}{1-n} \log(\mathrm{Tr}\rho_{A \tilde{A}}^n)=-  \frac{1}{1-n} \mathrm{Tr} \left[ \log \bigg((C_{A\tilde{A}}+1/2)^n-(C_{A\tilde{A}}-1/2)^n \bigg)\right] 
\end{equation}
For more details on this formula, see the review in appendix \ref{app:ReFn}. 

 \subsection{Gaussian wavefunction method for R\'enyi reflected entropy}
 There is an alternative way to compute the reflected entropy and its R\'enyi generalization in free scalar theory. In this case, the reduced density matrix for the subregion is a Gaussian form. After the canonical purification, one can easily show the purified state is also a Gaussian form. So, we can use almost the same algorithm for the R\'enyi EoP to calculate the reflected entropy. The only difference is that we don't need to vary the parameters to get the minimal entropy.  The task is that we need to solve these parameters from the equation demanded by the canonical purification.
 \begin{equation}
    \rho_{AB} = \rho_{\tilde{A} \tilde{B}}
 \end{equation}
 It will impose that 
 \begin{equation}
     L= J, \qquad  K = K^{T}
 \end{equation}
 Remember that we also have the constraints \eqref{purifiedConstraint}. With these equations, we can solve the matrix $K$. 
 
\begin{figure}
    \centering
    \begin{subfigure}[b]{0.48\textwidth}
        \centering
        \includegraphics[width=\textwidth]{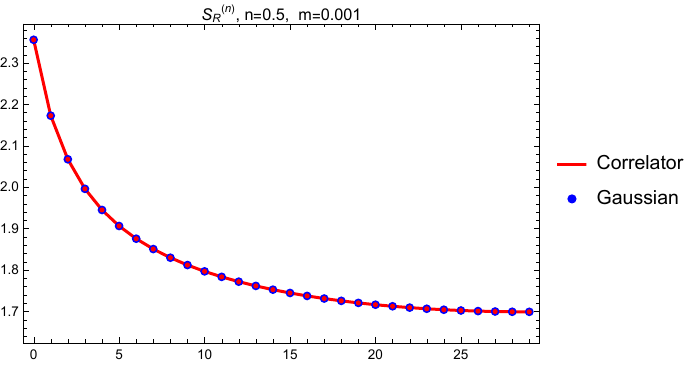}
    \end{subfigure}
    \hfill   % 分隔符
    \begin{subfigure}[b]{0.48\textwidth}
        \centering
        \includegraphics[width=\textwidth]{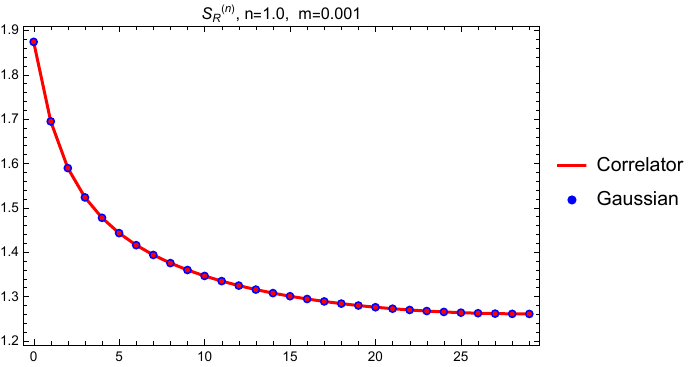}
    \end{subfigure} \\
    \begin{subfigure}[b]{0.49\textwidth}
        \centering
        \includegraphics[width=\textwidth]{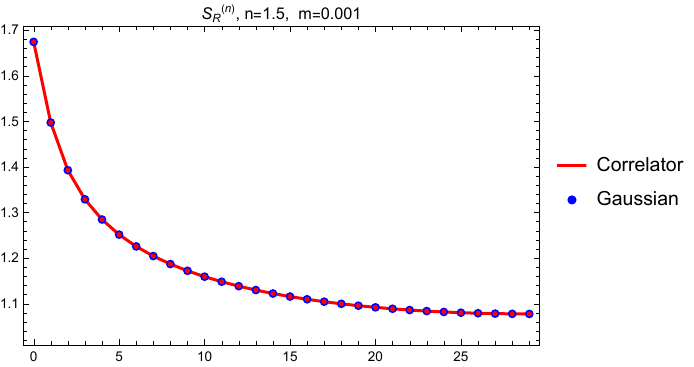}
    \end{subfigure}
    \begin{subfigure}[b]{0.49\textwidth}
        \centering
        \includegraphics[width=\textwidth]{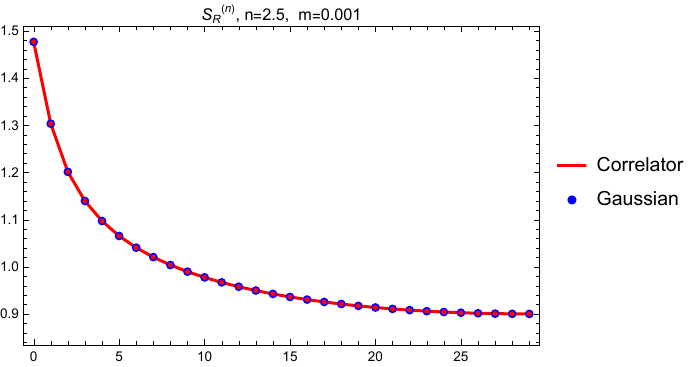}
    \end{subfigure}
    \caption{ The results of Renyi reflected entropy obtained from  correlator method (red) and Gaussian wavefunction method (blue) for $n=0.5, 1, 1.5$ and $2.5$, respectively. We connect the data of correlator method by line for comparison purpose. The total size is $N=60$, mass $m$ is $0.001$, and the size for subsystem is $|A| = |B| =1$. Due to the analytical equivalence of two methods, they match very well. }
     \label{two-methods}
\end{figure}

 Let us consider the simplest case as an example. Let the subsystem size be $|A| = |B| =1$, then we can parameterize matrix $K$ as
 \begin{equation}
    K=\begin{pmatrix}
        x & y  \\
        y & z \\
    \end{pmatrix} 
\end{equation}
Although the symmetry discussed in \cite{Takayanagi:2018sbw}  for EoP is broken for the reflected entropy, we can still assume $x>0$ by similar arguments. While we want to emphasize that the above equations are just necessary conditions. In this A1B1 case, we found two branches of solutions. One satisfies $x \ge y$, the other satisfies $x < y$. These two branches give different reflected entropy. Only the $x \ge y$ branch gives the right results. And it shows the right decreasing behavior when the distance of subregions increases. We also compare the result of this branch with the one obtained from correlator method, they match with each other very well, see Fig. \ref{two-methods} as an illustration.  Of course, we can also fix the right branch by a more careful analysis. While, we won't do this, since it need to solve the equation to determine the parameters.  We will adopt the correlator method in the following section which is more convenient.  

 \section{Numerical results for R\'enyi EoP and R\'enyi reflected entropy} \label{sec:NumericalResult}
 In this section, we present the numerical results of R\'enyi EoP and R\'enyi reflected entropy. We will compare R\'enyi EoP and half R\'enyi reflected entropy
 in different subsystem configurations. We will present two situations where the system size of AB is $|A| = |B| =1 $ and $  |A| = 1, |B| =2 $, respectively. We will also change separating distance $d$ of AB. For more general subsystem sizes,  we give a brief comment on the computational difficulties.

 \subsection{R\'enyi EoP and R\'enyi reflected entropy:  $|A| = |B| =1 $ } 
 Here, we present our numerical results for R\'enyi EoP and R\'enyi reflected entropy. The simplest subsystem size is $|A| = |B| =1 $. For the size of entire system, we will choose the same one as in previous literature \cite{Takayanagi:2018sbw},
 \ie $N = 60$. The $K$ matrix in the purified state is described by two parameters:
 \begin{equation}
    K=\begin{pmatrix}
        1 & x  \\
        y & 1 \\
    \end{pmatrix} 
 \end{equation}
 where this form already used the symmetry transformation discussed in Section 3.4 of \cite{Takayanagi:2018sbw}  to reduce the parameters. Besides that, since system size of $A$ and $B$ are the same, there is an extra $Z_{2}$ symmetry. It was shown that $x=y$ is always satisfied if we need the minimal entanglement of purification, so is R\'enyi EoP.  With the $K$ matrix parameterized, the $L$ matrix is obtained by:
\begin{equation}
      L^{-1}   =  K^{-1} \bigg(K_0 L_0^{-1} K_0^T \bigg)(K^{T})^{-1}
\end{equation}
and the matrix $J =J_{0}$ is invariant. 

For the R\'enyi reflected entropy, the story is much simpler. We just need to compute the $W$ matrix in original global pure state and its inverse. Then, restricting the index to subregion $A \cup B$ to obtain $X$ and $P$. Therefore, $G^{\Phi}$ and $G^{\Pi}$ are constructed. Taking the submatrix of $G^{\Phi}$ and $G^{\Pi}$ with support in 
$A \cup \tilde{A}$, the matrix $C_{A\tilde{A}}$ is obtained. Finally, by using formula \eqref{RenyiReF}, we get the R\'enyi reflected entropy.

\begin{figure}
     \centering
     \begin{subfigure}[b]{0.45\textwidth}
         \centering
         \includegraphics[width=\textwidth]{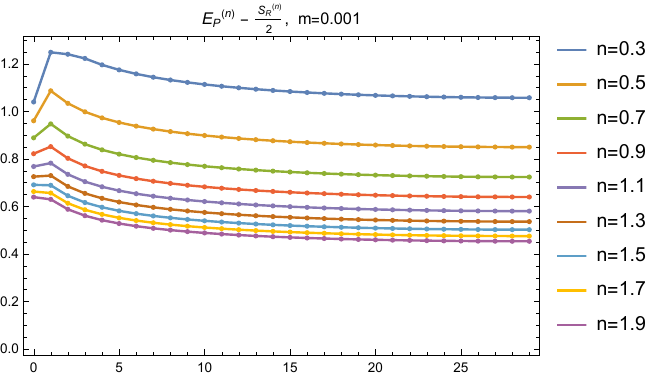}
     \end{subfigure}
     % 分隔符
     \hfill
     \begin{subfigure}[b]{0.45\textwidth}
         \centering
         \includegraphics[width=\textwidth]{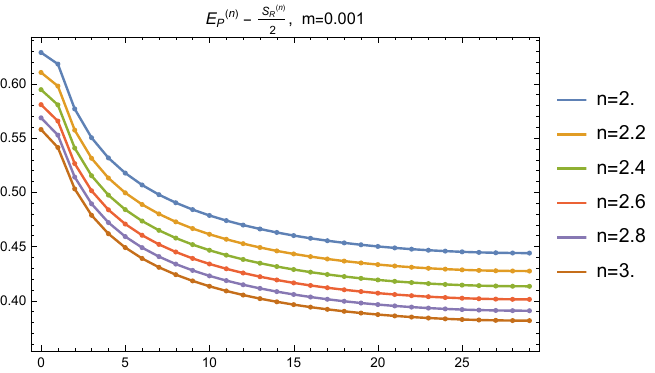}
     \end{subfigure}
      \begin{subfigure}[b]{0.45\textwidth}
         \centering
         \includegraphics[width=\textwidth]{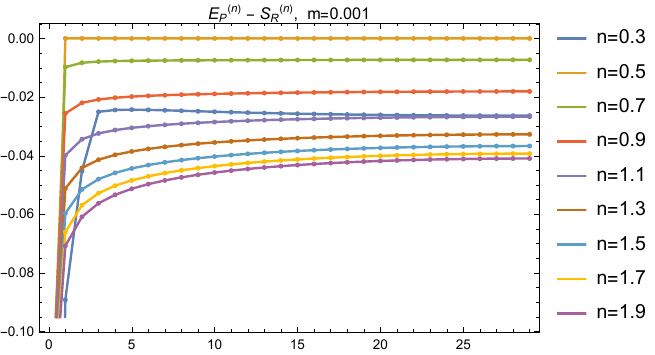}
     \end{subfigure}
     % 分隔符
     \hfill
     \begin{subfigure}[b]{0.45\textwidth}
         \centering
         \includegraphics[width=\textwidth]{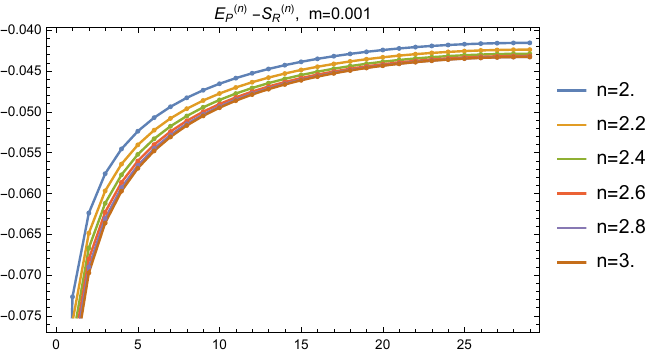}
     \end{subfigure}
     \caption{ The difference between R\'enyi Entanglement of Purification ($E_{P}^{(n)}$) and half R\'enyi reflected entropy with mass  $m=0.001$. (Bottom): The difference R\'enyi of Entanglement of Purification ($E_{P}^{(n)}$)  and R\'enyi reflected entropy with mass $m=0.001$. (Left):  for R\'enyi index $0< n < 2 $. (Right): for R\'enyi index $2 \le n \le 3 $. The horizontal axis is the distance between subregion $A$ and $B$ from zero to half length of the chain.}
     \label{fig:renyiEoP11}
 \end{figure}

 We plot the behavior of R\'enyi EoP and half R\'enyi reflected entropy with respect to different subregion separations, see the top of Fig. \ref{fig:renyiEoP11}. Since we choose the periodic boundary condition, we only need to plot the distance of the subregions up to half length of their complement region.
 Here, we also present the results for $E_p -S_{R}^{n}$, see the bottom of Fig. \ref{fig:renyiEoP11}. As one can see, $E_p$ is always smaller than $S_{R}^{n}$, that is because 
 the canonical purification is also a purification candidates for $E_p$ and the definition of $E_p$ is taking the minimal one. While, we should point out that 
 here the $E_p$  is computed by Gaussian ansatz. Even in this way, the parameters space is too large, so we take the same strategy in \cite{Takayanagi:2018sbw} 
 , we will choose the size of auxiliary system is the same as originals subsystem, this is called minimal (size) ansatz. The top part of Fig \ref{fig:renyiEoP11} shows $E_p$ is large than $S_{R}^{(n)}/2$ which is a more nontrivial result. Besides the results for R\'enyi index $n \in (0,2)$, we also show the R\'enyi index $n > 2$. For simplicity, we give the result for $n \in (2,3)$. Unlike the situation of RTNs, here we see the inequality
 \begin{equation}
    E_p^{(n)} \ge \frac{S_{R}^{(n)}}{2}
 \end{equation}
  holds for generic $n$. Even though, the free theory is not a holographic CFTs which has a semi-classical bulk dual, it does share some common features with holographic CFTs. So, inequality may hint it also holds for holographic CFTs. And in holographic CFTs, there may be a strong version for this, one expected that $E_p =\frac{S_{R}}{2} $. For the free scalar, only the inequality holds. 
  
  We also notice that the difference:
  \begin{equation}
     \Delta^{(n)} (A:B) = E_p^{(n)} -  \frac{S_{R}^{(n)}}{2}
  \end{equation} 
 exhibits nice decreasing behavior when the separation distance become larger. And it also decreases when the Renyi index gets larger which is also a nontrivial behavior,  since the decreasing behaviors of $ E_p^{(n)}$ and $ \frac{S_{R}^{(n)}}{2}$ do not guarantee decreasing behavior of $\Delta^{(n)} (A:B)$.  Note that, at the top of Fig \ref{fig:renyiEoP11}, it is increased from the first data point to the second for some R\'enyi $n$. While, this anomaly is because the separation distance changed from zero to one. When the distance is zero, the two subregions are connected. Thus, when the distance becomes one, the correlation between subregions get enhanced. In fact, this phenomenon has already been noticed in \cite{Takayanagi:2018sbw} for EoP (Note that our convention is little bit different from them, our $d$ corresponds to $d+1$ in their paper). We may conjecture that $\Delta^{(n)}(A:B)$ is also a information measure for certain class models. For $E_p^{(n)} -  {S_{R}^{(n)}} $, as shown in the bottom-left part of  \ref{fig:renyiEoP11}, some curves will cross over the others.

 \subsection{R\'enyi EoP and R\'enyi reflected entropy:  $|A| = 1, |B| =2 $  }
  \begin{figure}[t]
     \centering
     \begin{subfigure}[b]{0.46\textwidth}
         \centering
         \includegraphics[width=\textwidth]{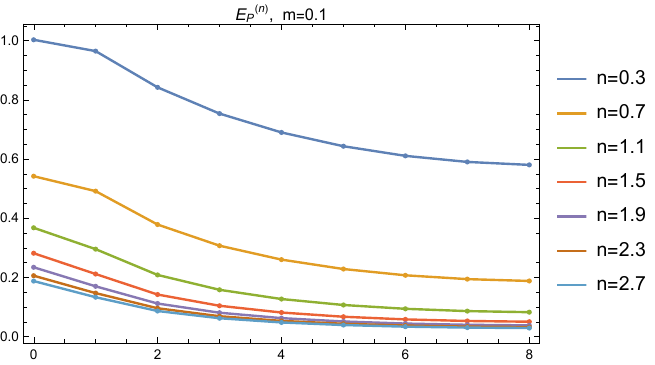}
     \end{subfigure}
     % 分隔符
     \hfill
     \begin{subfigure}[b]{0.46\textwidth}
         \centering
         \includegraphics[width=\textwidth]{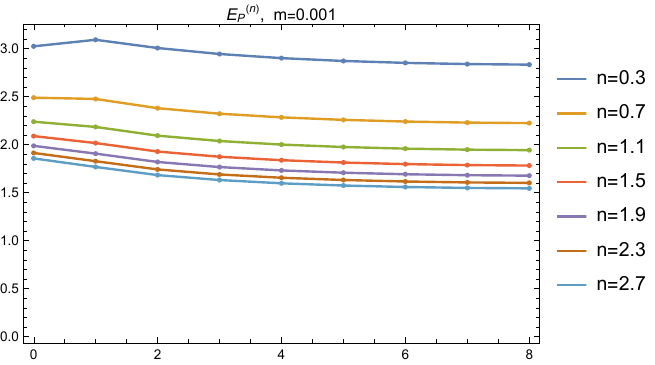}
     \end{subfigure}
     \caption{ $E_{P}^{(n)}$ for $|A|=1, |B|= 2$ case with different mass $m=0.1$  (left) and  $m= 0.001$ (right).  The horizontal axis is the distance between subregion $A$ and $B$. }
     \label{A1B2VaryMass}
 \end{figure}
 The above setup only contains two sites. We can make the subsystem a little larger, \ie let $A$ contains one site and $B$ contain two sites. In this case, we need four parameters to describe the purified state of EoP, and the matrix $K$ is:
    \begin{equation}
        K=\begin{pmatrix}
            1 & 0 & x \\
            y & 1 & 0 \\
            z & w & 1
        \end{pmatrix}
    \end{equation}  
where the symmetry transformation was used again. While due to the increasing computation complexity, we will consider a small system in this case and set the total size $N=20$. As for the mass parameter, we will focus on larger mass case, say $m=0.1$. This is because the decay behavior is not obvious for smaller mass. See Fig. \ref{A1B2VaryMass} for details.

\begin{figure}[t] 
    \centering
    \begin{subfigure}[b]{0.45\textwidth}
        \centering
        \includegraphics[width=\textwidth]{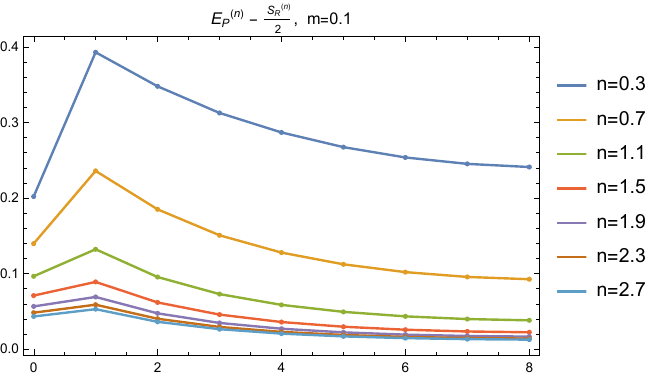}
    \end{subfigure}
    % 分隔符
    \hfill
    \begin{subfigure}[b]{0.45\textwidth}
        \centering
        \includegraphics[width=\textwidth]{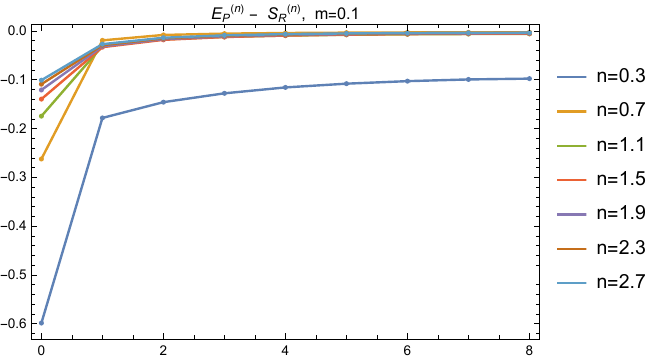}
    \end{subfigure}
    \caption{ Results of R\'enyi EoP and R\'enyi reflected entropy for $|A|=1, |B|= 2$ case, m=0.1. (Left):  R\'enyi EoP minus half R\'enyi reflected entropy. (Right): R\'enyi EoP minus R\'enyi reflected entropy.  The horizontal axis is the distance between subregion $A$ and $B$.  }
    \label{A1B2}
\end{figure}

For the R\'enyi reflected entropy, all procedures are the same as those in $|A| = 1, |B| =1 $ case. There is not much complexity involved, and one just needs to modify the global parameters. The combined results  of R\'enyi EoP and R\'enyi reflected entropy are presented in Fig. \ref{A1B2}.

 \subsection{ More general subsystem sizes: a brief comment }
 Consider that the subsystem $A$ and $B$ both contain two sites, then the $K$ matrix is parameterized as
 \begin{equation}
    K=\left(\begin{array}{cccc}
    1 & 0 & x & y \\
    0 & 1 & z & w \\
    x^{\prime} & y^{\prime} & 1 & 0 \\
    z^{\prime} & w^{\prime} & 0 & 1
    \end{array}\right)
    \end{equation}
Again, due to the symmetry between subsystem $A$ and $B$, we can check that numerically the minimal point  still satisfies $x =x^{\prime}, y = y^{\prime}, z= z^{\prime}, w= w^{\prime}$. So, we just need to minimize the EoP with respect to four parameters effectively. While, in this case, the matrix $\hat{\Lambda}$ discussed around \eqref{EoPLambda} becomes larger. It is $4 \times 4$ matrix now, so one can still have an analytical formula for $\hat{\lambda}_{i}(x,y,z,w)$ before the minimization step. Even though this is convenient for minimization, this formula in the code is tremendously complicated and makes the code run very slowly. 

Furthermore, for larger subsystems, not only does the number of parameters increase, but also the degree of the polynomial in the eigen equation will be greater than 4. Thus, according to Galois theory, in general, there is no closed formula available.  In this case, we need to assign the value for parameters in $K$ based on some observations and guesses. Then,  more sophisticated minimization algorithms are needed, for example one can design the simulated annealing algorithm for this R\'enyi EoP from the beginning to avoid using analytical expression in the middle step.
 
 \section{Summary and Discussions} \label{sec:Summary}
 In summary, we presented a pedagogical review on the Gaussian method of entanglement of purification in free  scalar and the correlator method for reflected entropy. With these general strategies in hand,  we presented the numerical results of R\'enyi entanglement of purification and  R\'enyi reflected   entropy in free scalar theory. We particularly focused on the R\'enyi index in region $0<n<2$. We have compared the entanglement of purification and half reflected entropy. The results show the inequality $E_{P}^{(n)}- S_{R}^{(n)}/2 \ge 0 $ holds in  $0<n<2$. For completeness, we also illustrated this inequality of $n>2$ (for illustration we just plot $2 \le n<3$).
 
 We would like to emphasize the inequality in region $0<n<2$ is crucial for the equivalence of holographic EoP and half holographic reflected entropy.  We also gave an alternative method for R\'enyi reflected entropy, the results are consistent with correlator method. The consistency of two methods also plays as a cross-check for our result. Also, for R\'enyi EoP in the $n=1$ case, both our analytical and  numerical results reduce to the results in \cite{Takayanagi:2018sbw}.
 
From the field theory perspective, we see the EoP and reflected entropy are two different concepts. One is the minimal purification, and the other is canonical purification.  Our calculations supported the inequality $ E_{P}^{(n) } \ge S_{R}^{(n)}/2$, thus make the holographic purification entropy conjecture more reliable. While, it is still an open question that how the holographic conditions remove their difference.

 \subsection{Mutual information and Markov gap}
        \begin{figure}[t]
               \centering
               \begin{subfigure}[b]{0.46\textwidth}
                   \centering
                   \includegraphics[width=\textwidth]{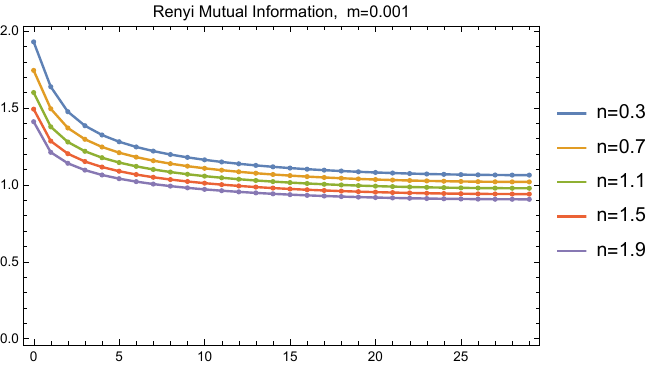}
               \end{subfigure}
               % 分隔符
               \hfill
               \begin{subfigure}[b]{0.46\textwidth}
                   \centering
                   \includegraphics[width=\textwidth]{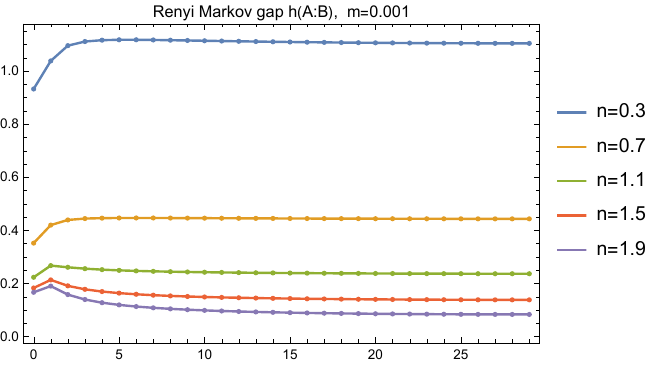}
               \end{subfigure}
               \caption{R\'enyi mutual information and R\'enyi Markov gap for $|A|=1, |B|= 1$ case, m=0.001, N=60. The horizontal axis is the distance between subregion $A$ and $B$. }
               \label{MIandMarkov}
       \end{figure}
 Here, we discuss some nice  properties of information measure to show the validity of our result.  Since in general, the reflected entropy satisfies the following inequality  \cite{Dutta:2019gen}:
 \begin{equation}
     \min\{2S_A,2S_B\}\geq S_R(A:B)\geq I(A:B).
 \end{equation}
 Let us focus on the right part as an illustration. The definition for Markov gap is given by
\begin{equation}
    h(A: B)=S_R(A: B)-I(A: B) ,
\end{equation}
 which is related to the fidelity of the Markov recovery map  \cite{Hayden:2021gno}.  We presented the numerical results for R\'enyi mutual information and R\'enyi Markov gap, see Fig. \ref{MIandMarkov}. It shows that the R\'enyi Markov is always positive as expected. For EoP, it was shown to satisfy the below equality \cite{Terhal:2002riz}:
 \begin{equation}
     \min\{ S_A, S_B\} \geq E_{P}(A: B) \geq \frac{1}{2} I(A: B)
 \end{equation}
Since we have demonstrated that $E_{P}^{(n)} \ge \frac{1}{2} S_{R}^{(n)}$, it follows that,  in the context of our numerical analysis, the positivity of the R\'enyi Markov gap entails the validity of the relation  $E_{P}^{(n)} \ge \frac{1}{2} I^{(n)} $.

  \subsection{Counter-examples}
   \begin{figure}[t]
      \centering
      \begin{subfigure}[b]{0.46\textwidth}
          \centering
          \includegraphics[width=\textwidth]{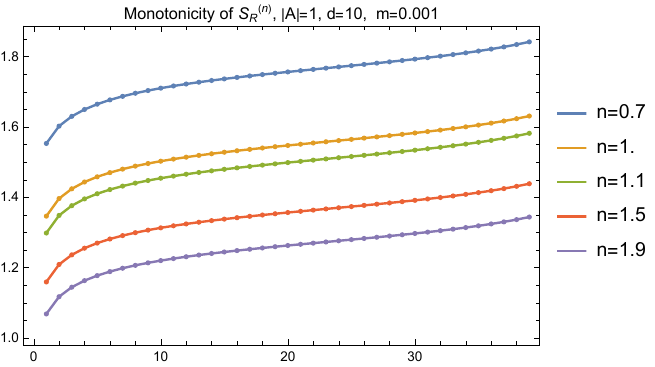}
      \end{subfigure}
      \hfill  % 分隔符
      \begin{subfigure}[b]{0.46\textwidth}
          \centering
          \includegraphics[width=\textwidth]{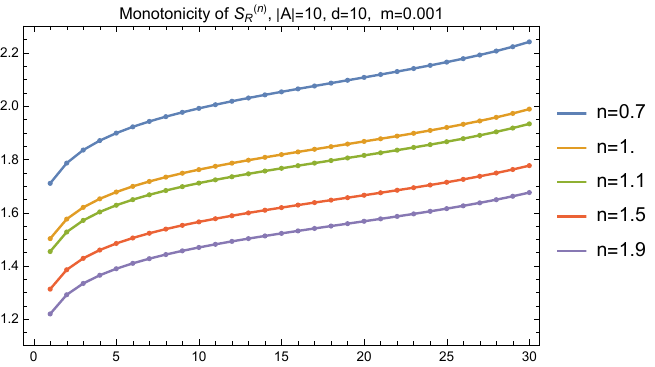}
      \end{subfigure}
      \caption{Monotonicity of R\'enyi reflected entropy.  (Right:) $|A|=10$.  The parameters are  $m=0.001, N=60, d=10$. (Left:) $|A|=1$.The horizontal axis is the size of $B$ ranged from $1$ to $N- |A| - 2d $. The R\'enyi indices are choosed as $n=0.7, 1.0, 1.1, 1.5, 1.9$ . }
      \label{Monotonicity}
  \end{figure}
   In \cite{Hayden:2023yij}, they constructed an explicit model consisting of two qutrits and a qubit, in which they showed that the monotonicity of (R\'enyi) reflected entropy is violated, they stated that this violation happens even at the classical level.  Similar violations also happen in its higher-dimensional Hilbert space extension \cite{Basak:2023uix}. Thus, in generic aspect, the (R\'enyi) reflected entropy is not a correlation measure.
   
   While their model is a very simple discrete toy model. For physical states in the continuum theories, the  reflected entropy still has the opportunity to be a good correlation measure. In holography context, Dutta and Faulkner already showed the monotonicity  holds for the integer R\'enyi index $n \ge 2$ based on entanglement wedge nesting \cite{Dutta:2019gen}. In \cite{Bueno:2020fle}, Bueno and Casini  also demonstrated that the monotonicity of reflected entropy is satisfied in free scalar and free fermion.
   
   The model we considered in this paper is indeed free scalar. We can also verify the general R\'enyi version of monotonicity
   \begin{equation}
      S^{(n)}_{R}(A: BC) \ge   S^{(n)}_{R}(A: B)
   \end{equation}
    including the reflected entropy itself by using our numerical algorithm. The results are presented in Fig. \ref{Monotonicity}, where we set the total size of system is $N=60$, we fix the distance of subsystem $A$ and $B$ as $d =10$. In left of Fig. \ref{Monotonicity}, we let the size of $A$ be 1, we change the size of $B$ from $1$ to $N-|A|- 2 d $. Since the correlator method for reflected entropy can be easier extended to larger subsystems, we also present the result of $A=10$  in the right of Fig. \ref{Monotonicity}. As these results show, these R\'enyi reflected entropies (including the reflected entropy itself ) show a nice monotonicity.
    
    We also noticed that in \cite{Couch:2023pav}, an explicit example which violate the inequality
    \begin{equation}
        E_P(A:B)\geq\frac{1}{2}S_R(A:B)
    \end{equation}
    was constructed. While, as they stated that it doesn't preclude the possibility for certain restricted states. In particular, for  states are holographic or close to holographic, it is expected this inequality is valid

   \subsection{The holographic conditions}
    In general, from a bottom - up view, we don't know the sufficient and necessary conditions for holographic CFT. Here, by holographic CFT, we mean it has a semi - classical gravity dual. However, to be a holographic CFT, we know that it needs to satisfy some necessary conditions, such as:
    \begin{itemize}
        \item conformal symmetry
        \item large central charge
        \item strong interactions
    \end{itemize}
    If one considers the dual between string theory and CFTs, some conditions listed above would be unnecessary. For example, the string theory on AdS$_{3} \times S^{3} \times T^{4}$  background  with pure NS - NS flux at minimal tension $k = 1$ is dual to symmetric product orbifold CFT which is free and the central charge is $c = 6 w$ \cite{Gaberdiel:2018rqv, Eberhardt:2018ouy, Eberhardt:2019ywk, Eberhardt:2020bgq}, and $w$ is the winding number. Although the theory is free and our strategy can be applied in principle in this case, we don't know how to compute the entanglement entropy in string theory in general. See \cite{Wong:2025kpz} for some attempts, nevertheless the entanglement of purification and reflected entropy.
    
    In our context, we have considered the free scalar theory, which satisfies the first condition \footnote{More precisely, the conformal symmetry is broken when the model is discretized. However, many properties are still inherited in the lattice model. For example, the well - known single interval  entropy formula  $S_{A}  =  \frac{c}{3} \log \frac{\ell}{\epsilon}$  also holds for the lattice model \cite{Holzhey:1994we,Vidal:2002rm,Calabrese:2004eu}. Also the modular Hamiltonian for the lattice model are well described by its analytical formula in continuous in some cases \cite{Kokail2021}. }. For the second condition, it is trivial for free scalar, since there are no interactions (the third condition), one can simply copy the original free scalar many times so that the central charge is large. To be holographic, the third condition is crucial. Unfortunately, our strategies for free scalars can not be migrated to these theories. They highly depend on the conditions of the free theory or Wick contraction. A more general treatment for holographic models seems elusive, so a more practical scheme to make progress would be focusing on the Random Tensor Networks (RTNs) which is a toy model for holography \cite{Hayden:2016cfa}. In \cite{Akers:2023obn}, their results on RTNs are valid only for integer Rényi index with $n \ge 2$.  The reason is that they need the replica trick, especially the lemma 1 at section II of their paper, which is valid for integer  $n \ge 2$. Therefore,  an open question is: Can we overcome this constraint in RTNs so that we will have the equivalence between entanglement of purification and half reflected entropy in a more holographic - like model.
    
    To understand how these holographic conditions work, a simpler question is to study how they make the reflected entropy a good measure. Since the entanglement of purification is a good information measure in general, the validity of the reflected entropy as a measure is the sufficient condition for the equivalence.

   \subsection{Remarks on other information measures}  
  The purification entropy is a kind of measure for correlation measures of the mixed states. One may wonder are there more information measures of the mixed state which have a clear geometry picture in the gravity side?  In fact, by doing the partial transposition of state $\rho_{AB}$ \cite{Peres:1996dw}:
  \begin{equation}
      \left\langle i_A, j_B\right| \rho_{AB}^{T_A}\left|k_A, l_B\right\rangle \equiv\left\langle k_A, j_B\right| \rho_{AB}\left|i_A, l_B\right\rangle ,
  \end{equation}
   where $i,j, k,l$ are the label for basses. Then, the logarithmic negativity \cite{Vidal:2002zz}:
   \begin{equation}
       \mathcal{E}(\rho):=\log \left\|\rho^{T_{A}}_{AB}\right\|_1
   \end{equation}
   was proposed, where $\|X \|_{1} = \mathrm{Tr} \sqrt{X X^{\dagger}}$ is the matrix trace norm. Even though the logarithmic negativity lacks the convex property, its monotonicity, which is crucial as an information measure, was proved in \cite {Plenio:2005cwa}. An advantage of the logarithmic negativity is its computability. Unlike the Entanglement of Purification that requires minimization procedures, it can be determined solely from the spectrum of $\rho_{AB}^{T_{A}}$. More simply, one can compute it using the replica trick. A systematic method to calculate the logarithmic negativity in QFTs was developed  in \cite{Calabrese:2012ew, Calabrese:2012nk}. For spherical entangling surface, it was shown that the logarithmic negativity $ \mathcal{E}$ is equivalent to entanglement wedge cross section with a bulk quantum correction term \cite{Kudler-Flam:2018qjo}:
  \begin{equation}
      \mathcal{E}=\mathcal{X}_d  E_W(A: B)+\mathcal{E}_{b u l k}
  \end{equation}
  where $\mathcal{X}_d$ is a constant depending on the dimension $d$ of spacetime and  $\mathcal{E}_{b u l k}$ is the bulk logarithmic negativity.
  
   Based on $\rho^{T_{A}}_{AB}$, one can also define the odd entanglement entropy (OEE) \cite{Tamaoka:2018ned}:
   \begin{equation}
       S_o\left(\rho_{A B}\right) \equiv \lim _{n_o \rightarrow 1} S_o^{\left(n_o\right)}\left(\rho_{A B}\right), \;\ S_o^{\left(n_o\right)}\left(\rho_{A B}\right) \equiv \frac{1}{1-n_o}\left[\operatorname{Tr}\left(\rho_{A B}^{T_{A}}\right)^{n_o}-1\right],
   \end{equation}
    where $n_{o}$ is the analytical continuation from an odd integer. For the vacuum state and the thermal state of CFT$_{2}$, the odd entanglement entropy $S_o\left(\rho_{A B}\right)$ was shown  to be dual to entanglement wedge cross section by \cite{Tamaoka:2018ned}:
  \begin{equation}
      S_o\left(\rho_{A B}\right)-S\left(\rho_{A B}\right)=E_W\left(A:B\right), 
  \end{equation}
  In general, $ S_o\left(\rho_{A B}\right)-S\left(\rho_{A B}\right)$ can be negative. While, $E_{W}(A:B)$ is always non-negative. This implies the holographic conditions also play non-trivial role for the above equality.  
   
  The partial entanglement entropy (PEE) is defined as \cite{Vidal:2014aal}:
   \begin{equation}
       s_A\left(A_i\right)=\int_{A_i} f_A(\mathbf{x}) d x^{d-1}.
   \end{equation}
   where $f_A(\mathbf{x})$ is the entanglement contour giving the entanglement contribution of each point, and $A_{i}$ is a subset of subregion $A$. Now, let the subregion be $A \cup B$, and require the balanced conditions as follows \cite{Wen:2021qgx}:
   \begin{equation}
       s_{A A^{\prime}}(A)=s_{B B^{\prime}}(B), \quad s_{A A^{\prime}}\left(A^{\prime}\right)=s_{B B^{\prime}}\left(B^{\prime}\right)
   \end{equation} 
   where $A', B'$ are the auxiliary regions to purify $\rho_{AB}$. The Balanced Partial Entanglement (BPE) is the  partial entanglement entropy $s_{AA'} (A)$ under the above balanced conditions. The BPE was shown dual to entanglement wedge cross section for the holographic purification as well \cite{Wen:2021qgx}. 
  
    In \cite{Jiang:2024akz}, a new kind of way for purification by subtracting the undetectable regions in CFT$_{2}$ was proposed. By using their subtraction proposal, they confirmed the corresponding purified entropy is dual to the EWCS. To reproduce their results, one just needs to do the replica trick and conformal transformation on the subtracted geometry, no details regarding purification and conformal block data are needed. 
   
   All these measures are related to entanglement wedge cross section in holography, and it would be nice to clarify their relations in the future.
   
   \subsection{Some possible ways to extend our current results}
  In this subsection, we would like to give a road-map to the audiences who want to extend our results in the future. 
  \begin{itemize}
      \item \textbf{Generalize to free fermion:} The setup of this paper is free scalar, while the methods we applied mainly depend on the free theory structure, for example the Wick theorem, so we believe that our method can be extended to the free fermion case, and we expect that these inequalities will still hold in the free fermion case. 
      
      \item \textbf{Consider the free symmetric product orbifold CFT:} The string theory on AdS$_{3} \times S^{3} \times T^{4}$  background with pure NS-NS flux at minimal tension $k = 1$ is precisely dual to symmetric product orbifold CFT which is free. Although it is not clear how to compute the entropy-related quantities in the string side of this duality, it will be quite inspiring to extend our results to the field side in the future.
      
      \item  \textbf{Consider the other boundary conditions and excited states:}  In this paper, we considered the periodic boundary condition, the global pure state is chosen as the ground state. It would be nice to explore the effects of the other boundary conditions and  the excited states systematically.
      
      \item \textbf{Extend to larger subsystem sizes: } For the (R\'enyi) EoP,  both our results and the results in \cite{Takayanagi:2018sbw} are restricted to small subsystem sizes. So one may wonder how the finite size effects impact our results. Although we expect that they may not change the final statements, It would be nice to explore them in the future to put these statements on a more solid footing. In general, we need to minimize an objective function that contains $2|A||B|$ parameters.  It is possible to develop a more sophisticated numerical code to extend the results to a larger system size. For example, one can try to use the simulated annealing algorithm or differential evolution algorithm to implement the minimization process. These are the purely numerical and global search minimization algorithms. It is necessary to avoid symbolic calculations from the very beginning, so as to apply this algorithm and greatly accelerate the calculation efficiency. For the very large subsystem sizes, we may expect that some parallel algorithms and super-computer servers are both needed. Once these large subsystem sizes cases are achieved, one can check the minimal Gaussian ansatz introduced in \cite{Takayanagi:2018sbw} in a broader range.
  \end{itemize} 
  Of course, there may be other possibilities to extend our research as we discussed in the former subsections. We believe the roads we listed above will be relatively straightforward and  practical. They will be insightful for interested readers.

\section*{Acknowledgments}
 L.C. would like to thank Tadashi Takayanagi and Shan-Ming Ruan for the inspiring discussions on this project. L.C. is supported by the Shuimu Tsinghua Scholar Program of Tsinghua University. L.C. is also grateful for the support from the Kavli Institute for Theoretical Sciences (KITS) at UCAS and for the hospitality provided by the Yukawa Institute for Theoretical Physics (YITP) at Kyoto University in the early stage of this work.

 \appendix
 
 \section{More details on correlator method for R\'enyi reflected entropy} \label{app:ReFn}
 Now, let us review the Bueno and Casini's method for the calculation of reflected entropy \cite{Bueno:2020fle}. Remember that the Hamiltonian of our model on the lattice is
 \begin{equation} \label{theHam}
     H=  \sum_{i=1}^N \frac{1}{2} \pi_i^2+\sum_{i, j=1}^N \frac{1}{2} \phi_i V_{i j} \phi_{j}
 \end{equation}
For later convenience, we give the explicit form of $V$ and $\sqrt{V}$ here:
 \begin{equation}
    V_{n n^{\prime}}=N^{-1} \sum_{k=1}^N\left[a^2 m^2+2(1-\cos (2 \pi k / N))\right] e^{2 \pi i k\left(n-n^{\prime}\right) / N} = \sum_{k} U_{n k} \mathcal{V}_{kk} ( U^{\dagger})_{k n^{\prime}}
\end{equation}
\begin{equation}
    \begin{aligned}    
       \sqrt{V} &=\sum_{k} U_{n k} \sqrt{\mathcal{V}_{kk}} ( U^{\dagger})_{k n^{\prime}}=\frac{1}{N} \sum_{k=1}^N \sqrt{a^2 m^2+2(1-\cos (2 \pi k / N))} e^{2 \pi i k\left(n-n^{\prime}\right) / N}
    \end{aligned}
\end{equation}
where the unitary matrix $U_{n k} ={1 \over \sqrt{N}} e^{ 2\pi i kn/N}$.

 The key point of Bueno and Casini's method is using the correlator to represent the reflected entropy. To compute the reflected entropy, the first step is extending the original Hilbert space $\mathcal{H}_{AB}$ of sub-systems $A,B$ with a double copy $\mathcal{H} = \mathcal{H}_{AB} \otimes \mathcal{H}_{\tilde{A}\tilde{B}}$. Then, the field operator $\phi_{i}$ and momentum operator $\pi_{j}$, $i,j = 1, 2, \cdots N_{AB}$, be extended by the following modular conjugation $J$
 \begin{equation}
     \tilde{\phi}_{i} = J \phi_{i} J, \qquad \tilde{\pi}_{i} = -J \pi_{i} J
 \end{equation}
 one can check these $4N_{AB}$ operators form the canonical commutation relations. In the following, we will simply write $N_{AB}$ as $N$ when there is no confusion. Now, let us arrange these operators into $\Phi$ and $\Pi$, s.t.
 \begin{equation}
     \begin{aligned}
         \Phi_{i} &= \phi_{i},  \quad \Phi_{i+N} &= \tilde{\phi}_{i} \\
         \Pi_{i} &= \pi_{i},  \quad \Pi_{i+N} &= \tilde{\pi}_{i} 
     \end{aligned}
 \end{equation}
 where $i=1,2, \cdots N $. We will focus on the following two correlators:
 \begin{equation}
     G_{ij}^{\Phi} = \langle \Omega | \Phi_{i} \Phi_{j} |\Omega \rangle, \quad G_{ij}^{\Pi} = \langle \Omega | \Pi_{i} \Pi_{j} |\Omega \rangle
 \end{equation}
 with the index $i= 1,2, \cdots, 2N$ in this case. In the following, we will show how to use the correlators 
 \begin{equation}
     X_{ij} = \langle \phi_{i} \phi_{j} \rangle,\quad  P_{ij} = \langle \pi_{i} \pi_{j} \rangle
 \end{equation}
 in the original Hilbert space to represent $ G_{ij}^{\Phi}$ and  $ G_{ij}^{\Pi} $.
 
 Since, we are considering the free theory, due to Wick theorem, the modular Hamiltonian for the subsystem will have the following quadratic form:
 \begin{equation}
     \mathcal{H}_{AB} = \sum_{j=1}^{N} \epsilon_{j} a_{j}^{\dagger} a_{j}
 \end{equation}
 Here, $\epsilon_{j}$ is the energy on the $j$ oscillator $a_{j}^{\dag} |0\rangle $. Then, the reduced density matrix is Gaussian:
 \begin{equation} \label{GaussianRho}
     \rho_{A B} =  \mathcal{N} \exp \big( - \sum_{j=1}^{N} \epsilon_{j} a_{j}^{\dagger} a_{j} \big), \quad  \mathcal{N} = \prod_{\ell=1}^{N} (1- e^{-\epsilon_{\ell}})
 \end{equation}
Then one can show that these two correlators have the following block-matrix representation.
 \begin{equation} \label{doubleCorrelator}
     \begin{aligned}
         & G^{\Phi}=\left(\begin{array}{cc}
             \alpha(2 n+1) \alpha^T & 2 \alpha \sqrt{n(n+1)} \alpha^T \\
             2 \alpha \sqrt{n(n+1)} \alpha^T & \alpha(2 n+1) \alpha^T
         \end{array}\right), \\
         & G^{\Pi}=\left(\begin{array}{cc}
             \frac{1}{4}\left(\alpha^{-1}\right)^T(2 n+1) \alpha^{-1} & -\frac{1}{2}\left(\alpha^{-1}\right)^T \sqrt{n(n+1)} \alpha^{-1} \\
             -\frac{1}{2}\left(\alpha^{-1}\right)^T \sqrt{n(n+1)} \alpha^{-1} & \frac{1}{4}\left(\alpha^{-1}\right)^T(2 n+1) \alpha^{-1}
         \end{array}\right) 
     \end{aligned}
 \end{equation}
 where $\alpha$ is a $N \times N$ coefficient matrix and is defined as 
 \begin{equation}
     \phi_i=\alpha_{i j}\left[a_j^{\dagger}+a_j\right] .
 \end{equation}
 And $n$ is a $N$ dimensional diagonal matrix, with diagonal elements are given by boson distribution:
 \begin{equation}
     \begin{aligned}
          n_{k k} &\equiv \left\langle a_k^{\dagger} a_k\right\rangle \\
          &= \mathrm{Tr} \bigg(\exp(- \sum_{j} \epsilon_{j} a_{j}^{\dag} a_{j}) a_{k}^{\dag} a_{k}\bigg) \prod_{\ell=1}^{N} (1- e^{-\epsilon_{\ell}}) \\
           &= \frac{1}{e^{\epsilon_k}-1}
     \end{aligned}
 \end{equation}
 On the other hand, we can use the canonical coordinates $\phi_{i}, \pi_{i}, i = 1,2, \cdots N$ to rewrite the reduced density matrix. Note that:
\begin{equation}
    [\phi_{i}, \pi_{j}] = i \delta_{ij}, \quad  [a_{i}, a_{j}^{\dag}] = \delta_{ij}
\end{equation}  
The Bogoliubov transformation is:
\begin{equation}
    \phi_{i} = \sum_{j} \alpha_{i j} \left[a_{j} + a_{j}^{\dag}\right], \quad \pi_{i} = i \sum_{j} \beta_{i j} \left[a_{j} - a_{j}^{\dag}\right], \quad \beta = -\frac{1}{2} (\alpha^{-1})^{T}
\end{equation}
Then, we can show:
\begin{equation}
    \begin{aligned}
        X_{ij} = \langle \phi_{i} \phi_{j} \rangle  
        = \sum_{k} \alpha_{i k} \bigg(2 \langle a_{k}^{\dag}a_{k}\rangle +1 \bigg) \alpha_{j k}  
         = \left(\alpha (2{n}+1)\alpha^{T} \right)_{ij}
    \end{aligned}
\end{equation}
Similarly, we have:
\begin{equation}
    P = \frac{1}{4} (\alpha^{-1})^{T} (2 {n} + 1) (\alpha^{-1})
\end{equation}
then, the corresponding correlators in the double copy Hilbert space are:
\begin{equation} \label{blockC}
    G^{\Phi}=\left(\begin{array}{cc}
        X & g(X P) X \\
        g(X P) X & X
    \end{array}\right), \quad G^{\Pi}=\left(\begin{array}{cc}
        P & -P g(X P) \\
        -P g(X P) & P
    \end{array}\right)
\end{equation}
where $G(O) = (O - 1/4)^{1/2} O^{-1/2}$. 

\subsection*{From the Gaussian ground state to determine $X$ and $P$}
In our discussion, the original pure state is the ground state which has a Gaussian structure. We can write this wave function explicitly from the free Hamiltonian (For original derivation see \cite{Bombelli1986}, for more details see \cite{Srednicki:1993im, Vidal:2002rm, Casini:2009sr}):
\begin{equation}
    \Psi_0\left[\phi_{i}, \phi_{\bar{i}}\right]=\mathcal{N}_0 \cdot \exp \left[-\frac{1}{2}\left(\phi_{i}, \phi_{\bar{i}}\right)\left(\begin{array}{cc}
        J_0 & K_0 \\
        K_0^T & L_0
    \end{array}\right)\left(\begin{array}{c}
        \phi_{i} \\
        \phi_{\bar{i}}
    \end{array}\right)\right]
    \label{GroundState}
\end{equation}
where $\phi_{i} $ means the components of $\phi$ with index in subregion $A B$, and $\bar{i}$ are the complement of $AB$. Now, the question is how we use this original pure state to determine the $(X,P)$  in correlator method. That is what is the relation between $ (J_{0}, K_{0}, L_{0})$ and $(X,P)$. 

Remember that all information of matrices $(J_{0}, K_{0},L_{0})$ comes from the potential matrix $V$ in the Hamiltonian, \ie
\begin{equation}
    V^{1/2} = \left(\begin{array}{cc}
        J_0 & K_0 \\
        K_0^T & L_0
    \end{array}\right)
\end{equation}
On the other hand, from the standard path integral method or canonical quantization method, we know the two point correlators of Hamiltonian \eqref{theHam} are totally determined by $V$ matrix, and have the following closed form:
\begin{equation}
    \begin{aligned}
        X_{i j} & =\left\langle\phi_i \phi_j\right\rangle=\frac{1}{2}\left(V^{-\frac{1}{2}}\right)_{i j} \\
        P_{i j} & =\left\langle\pi_i \pi_j\right\rangle=\frac{1}{2}\left(V^{\frac{1}{2}}\right)_{i j} =\frac{1}{2} \left(J_{0}\right)_{ij}
    \end{aligned}
\end{equation}

 \subsection*{Correlator method for the entanglement entropy }
 Before discussing the reflected entropy, it would be insightful to discuss entanglement entropy firstly. By using the Gaussian form \eqref{GaussianRho}, the entanglement entropy is 
 \begin{equation}
    S_{AB} = - \mathrm{Tr} \bigg(\rho_{AB} \log \rho _{AB}\bigg) = - \log \mathcal{N} + \sum_{l} \epsilon_{l} \langle a_{l}^{\dag} a_{l} \rangle
 \end{equation}
Substitute $\mathcal{N}$ and $n_{ll}$, it is simplified to 
 \begin{equation}
     \begin{aligned}
         S_{AB} &=\sum_l \left(-\log(1-e^{-\epsilon_l})+\frac{\epsilon_l e^{-\epsilon_l}}{1-e^{-\epsilon_l}}\right) \\
         &= \sum_l \bigg( \frac{e^{\epsilon_{l}}}{e^{\epsilon_{l}}-1} \log \frac{e^{\epsilon_{l}}}{e^{\epsilon_{l}}-1} -  \frac{1}{e^{\epsilon_{l}}-1} \log \frac{1}{e^{\epsilon_{l}}-1} \bigg)\\
         &=\mathrm{Tr}\left(( C_{AB} +1/2)\log( C_{AB} +1/2)-( C_{AB} -1/2)\log( C_{AB} -1/2)\right) 
     \end{aligned}
 \end{equation}
 where $ C_{AB} = \sqrt{G^{\Phi}_{AB} G^{\Pi}_{AB} } = \sqrt{X_{AB} P_{AB} }$ is an correlation matrix with index support in subregion $AB$. And the eigenvalue $ v_{k}$ of $ C_{AB} $ is related with $\epsilon_{k}$ by  
  \begin{equation}
     v_{k} =\frac{1}{2}(2 n_{kk} + 1) =\frac{1}{2} \coth(\epsilon_{k}/2) > \frac{1}{2}
 \end{equation}
 Similarly, we can generalize these formalism to R\'enyi entropy, Let us  first compute $\mathrm{Tr} \left( \rho_{AB}^{n}\right) $, 
 \begin{equation}
          \mathrm{Tr} \left( \rho_{AB}^{n}\right) = \frac{\prod_{l} (1 - e^{-\epsilon_l})^{n}}{\prod_{l} (1 - e^{-n\epsilon_l})} 
 \end{equation}
Then     
\begin{equation}
   \log \mathrm{Tr} \left( \rho_{AB}^{n}\right) = -\sum_{l} \log \frac{e^{n\epsilon_l} -1 }{(e^{\epsilon_l}-1)^{n}} = - \sum_{l} \log \bigg[ \Big(\frac{e^{\epsilon_l}}{e^{\epsilon_l}-1}\Big)^{n} - \Big( \frac{1}{e^{\epsilon}-1}\Big)^{n}  \bigg]
\end{equation}
 Thus, we have
 \begin{equation} \label{RenyiEE}
    S_{AB}^{(n)} = \frac{1}{1-n} \log(\mathrm{Tr}\rho_{AB}^n)=-  \frac{1}{1-n} \mathrm{Tr} \left[ \log \bigg(( C_{AB} +1/2)^n-( C_{AB} -1/2)^n \bigg)\right] 
\end{equation}

\subsection*{Correlator method for the reflected entropy}
 In this canonical purified system $AB \tilde{A} \tilde{B}$, the reflected entropy is just the entanglement entropy between $A\tilde{A}$ and $B \tilde{B}$. Thus, we can apply the above discussion for entanglement entropy $S_{AB}$ by replacing  $ C_{AB} $ as
 \begin{equation}
     C_{A\tilde{A}} = \sqrt{G^{\Phi}_{A\tilde{A}} G^{\Pi}_{A\tilde{A}}}
 \end{equation}
 This subregion correlation matrix $ C_{A\tilde{A}} $ can be easily constructed by restricting each sub-block of $G^{\Phi}$ and $G^{\Pi}$ in \eqref{blockC} to region $A$. Thus, in conclusion, we have
 \begin{equation}
     \begin{aligned}
         S_{R}(A : B)= S_{A \tilde{A}} &= \operatorname{tr}\left[\left( C_{A\tilde{A}}+1 / 2\right) \log \left(  C_{A\tilde{A}} + 1 / 2\right)-\left( C_{A\tilde{A}}-1 / 2\right) \log \left(  C_{A\tilde{A}}-1 / 2\right)\right] \\
     \end{aligned}
     \label{RefE}
 \end{equation}
Similarly, the R\'enyi reflected entropy is 
\begin{equation}
    S_{R}^{(n)}(A : B) = \frac{1}{1-n} \log(\mathrm{Tr}\rho_{A \tilde{A}}^n)=-  \frac{1}{1-n} \mathrm{Tr} \left[ \log \bigg((C_{A\tilde{A}}+1/2)^n-(C_{A\tilde{A}}-1/2)^n \bigg)\right] 
\end{equation}

\bibliographystyle{JHEP}
\bibliography{ref.bib}

\end{document}